\pdfoutput=1
\documentclass[11pt,letterpaper]{article}

\usepackage[margin=1in]{geometry}
\usepackage[T1]{fontenc}
\usepackage[utf8]{inputenc}
\usepackage{lmodern}
\usepackage{microtype}
\usepackage{amsmath}
\usepackage{amssymb}
\usepackage{bm}
\usepackage{booktabs}
\usepackage{array}
\usepackage{tabularx}
\usepackage{graphicx}
\usepackage[numbers,sort&compress]{natbib}
\usepackage{algorithm}
\usepackage{algpseudocode}
\usepackage{placeins}
\usepackage{xurl}
\usepackage[hidelinks]{hyperref}

\hypersetup{
    pdftitle={MoRF-AST: Calibrated Probabilistic Virtual Sensing for Structural Monitoring under Changing Operating Conditions},
    pdfauthor={Wingho Feng, Quanwang Li, Ming Zhong, Jingyu Yang, Chen Wang}
}
\title{MoRF-AST: Calibrated Probabilistic Virtual Sensing for Structural Monitoring under Changing Operating Conditions}
\author{
    Wingho Feng$^{1}$, Quanwang Li$^{1}$, Ming Zhong$^{2}$,\\
    Jingyu Yang$^{2}$, and Chen Wang$^{1,*}$\\[0.5em]
    \small $^{1}$Department of Civil Engineering, Tsinghua University, Beijing 100084, China\\
    \small $^{2}$Xiandai Investment Co., Ltd., Changsha 410004, China\\
    \small $^{*}$Corresponding author: \href{mailto:chwang@tsinghua.edu.cn}{chwang@tsinghua.edu.cn}
}
\date{August 2026}

\begin{document}
\maketitle

\begin{abstract}
Probabilistic full-field reconstruction provides uncertainty-aware response evidence for structural reliability assessment, yet inference from sparse and noisy measurements remains underdetermined.
Most existing methods overlook shifts between offline training and operational distributions. Under such shifts, posterior intervals may become miscalibrated, causing the reported uncertainty to lose its probabilistic meaning.
This study proposes Modal Residual Flow Matching with Context-Conditioned Affine Spread Transport (MoRF-AST) for calibrated structural virtual sensing under changing operating conditions. MoRF constructs an analytic Gaussian reference posterior in normalized modal coordinates and trains a conditional flow only on posterior-whitened residuals. At deployment, AST estimates response scale from historical measurements at installed sensors and uses gated, mean-preserving Bures--Wasserstein transport to adjust posterior spread. On a bridge-deck benchmark, MoRF achieves a posterior-mean normalized root-mean-square error (NRMSE) of 7.20\%, compared with 16.1\% and 17.9\% for two direct conditional flows. Across eight shifted traffic domains, AST reduces MoRF's cross-domain average coverage error from 0.0535 to 0.0236, a 55.9\% reduction, while preserving posterior-mean accuracy. The same transport does not improve the tested alternatives in aggregate, showing that calibration gains require its direction to match the base posterior's dispersion bias. MoRF-AST provides a data-efficient framework for probabilistic full-field reconstruction whose uncertainty remains interpretable under scale-dominated operational distribution shifts.
More broadly, this work highlights the need to calibrate uncertainty under changing operational distributions, thereby supporting trustworthy probabilistic modeling and reliability-informed decision-making in civil and infrastructure engineering.
\end{abstract}


\vspace{1em}

\section{Introduction}
\label{sec:introduction}

Large-scale engineering structures are instrumented at only a limited number of locations, whereas condition assessment, anomaly localization, and maintenance planning often require distributed displacements, strains, or internal forces. Access constraints, acquisition infrastructure, maintenance demands, and long-term reliability limit the spatial resolution that can be achieved by adding physical sensors. Virtual sensing therefore provides a practical link between sparse measurements and full-field structural response by reconstructing responses at uninstrumented locations \cite{kullaa2019bayesianvirtual,cao2023fullfield}. Deployment also exposes reconstruction models to operating conditions that evolve over time. Variations in loading, environmental conditions, and operating states can shift the structural response distribution away from the source domain \cite{sohn2007eov}. Virtual sensing must therefore close the spatial information gap while quantifying how strongly current sparse measurements constrain the full field under changing operating conditions. The resulting response distribution can serve as one reliable source of screening evidence for deciding whether a detailed, computationally intensive reliability analysis is warranted.

Sparse and noisy measurements generally do not uniquely determine a high-dimensional response field. A single reconstruction therefore conceals the multiple full-field states that remain compatible with the current measurements. We cast this task as learning a conditional posterior in a reduced representation. The posterior mean provides a point reconstruction, while its spread and dependence describe the conditional uncertainty left by measurement sparsity and noise. A reliable probabilistic reconstruction should combine low reconstruction error with marginal intervals whose empirical coverage agrees with their nominal levels, thereby avoiding overconfidence from undercoverage and excessively conservative intervals from overcoverage \cite{gundersen2021semiconditional,zou2023virtualsensing}.

Two challenges are central to the deployment setting considered here. First, direct conditional generative methods learn the full conditional state from finite paired sensor and full-field data. Under the retained observation model and known measurement noise, a Gaussian approximation already supplies an analytic posterior center and covariance. A direct conditional generator must nevertheless relearn this available geometry together with the non-Gaussian residual structure from finite paired data \cite{padmanabha2021conditional,karumuri2024amortized,dasgupta2025conditionalscore}. Second, posterior uncertainty depends on the source training distribution. When an offline dataset is deliberately balanced across prescribed regimes, its sampling weights become part of the learned source distribution. Without deployment adaptation, the model continues to use the source prior or conditional relationship induced by that design. The regime frequencies and response scales encountered on site may differ, so the source posterior can produce intervals that are too wide or too narrow \cite{qiu2023predictionsets}. This issue affects most probabilistic deep learning models that claim to provide uncertainty intervals, yet it is often overlooked. Retraining the same supervised posterior model would require new target-condition full fields, whereas the only additional target-domain information considered here consists of historical measurements from the installed sensors. The source model must therefore be retained while its posterior uncertainty is adjusted using those histories.

To address these limitations, we propose Modal Residual Flow Matching with Context-Conditioned Affine Spread Transport, termed MoRF-AST, which integrates source-domain posterior learning with deployment-time uncertainty adjustment. MoRF combines an analytic Gaussian reference with residual learning to construct a base posterior from the current measurement. AST uses historical sensor measurements from the current operating condition to adjust posterior spread while preserving its mean. This separates event-specific reconstruction from operating-condition spread adjustment.

There are three main contributions. First, MoRF learns a conditional residual distribution around an analytic Gaussian posterior. This improves data efficiency and provides shared posterior coordinates under a fixed sensing setup. Second, AST adapts posterior spread using histories from the installed sensors. It combines effective-scale estimation, truncation-aware gating, and mean-preserving Bures transport without target full fields or generator retraining. Third, we identify and verify that spread transport improves calibration only when its contraction or expansion agrees with the base posterior's dispersion bias. This explains why AST benefits MoRF but not the alternative base models across all shifted domains.

The structure of the paper is summarized as follows. Section 2 reviews probabilistic full-field reconstruction, conditional generative modeling, and deployment adaptation, and Section 3 presents MoRF-AST. Section 4 reports bridge-deck validation and calibration across operating domains, Section 5 examines design choices, ablations, deployment sensitivities, and recurrent operating histories, Section 6 discusses applicability and limitations, and Section 7 concludes the paper.
 \section{Related Works}
\label{sec:related-works}

This section reviews developments in structural virtual sensing and probabilistic full-field reconstruction, discusses generative posterior modeling for inverse problems, and then examines adaptation under operational distribution shift. This review identifies the remaining research gaps and outlines the objectives of the present study.

\subsection{Structural virtual sensing and probabilistic full-field reconstruction}
\label{sec:related-structural-virtual-sensing}

Structural virtual sensing uses a limited set of measured channels to reconstruct unmeasured displacements, accelerations, strains, and response fields at inaccessible or uninstrumented locations. Model-driven methods establish measurement-to-field relations through modal expansion, dynamic substructuring, reduced-order bases, and regularized inversion \cite{kullaa2016virtualsensing,cao2023fullfield,zhao2026physicsaligned}. Mechanics-based model updating can jointly infer structural parameters, unknown loads, and unmeasured responses \cite{nabiyan2021jointestimation}. State-space estimators use Kalman filtering and smoothing to fuse heterogeneous measurements, including asynchronous data acquired at different rates \cite{zhu2024multiratekalman}. Bayesian response expansion and Gaussian-process latent-force models further quantify predictive uncertainty by propagating uncertainties associated with loads, models, and measurements into conditional means, covariances, or credible intervals \cite{kullaa2019bayesianvirtual,zou2023virtualsensing}.

Data-driven methods learn nonlinear maps from monitored channels to unmeasured responses. MLPs and DNNs support sensor selection and response regression, while CNNs exploit temporal and cross-channel dependence for missing-response reconstruction \cite{pan2022dnnvirtualsensing,fan2021cnnvirtualsensing}. GANs, VAEs, and conditional diffusion models extend these mappings through generative response reconstruction, while latent-variable and diffusion formulations can produce stochastic response ensembles beyond a single deterministic estimate \cite{fan2023saganvirtualsensing,lee2026csvae,shu2025diffusionvirtualsensing}. Together, these developments broaden data-driven virtual sensing from point reconstruction toward stochastic response ensembles, complementing model-driven approaches to probabilistic full-field reconstruction. MoRF-AST builds on this progression by coupling conditional full-field posterior reconstruction with a dedicated deployment-time mechanism for adapting posterior spread under changing operating conditions.

\subsection{Generative posterior modeling for inverse problems}
\label{sec:related-generative-posterior-modeling}

Generative methods model posteriors for imaging, flow-field recovery, subsurface characterization, inverse elasticity, and structural response reconstruction. Given paired observations and targets, VAEs, conditional GANs, conditional normalizing flows, diffusion models, and flow-matching models can learn the conditional distribution directly and draw posterior samples for new observations \cite{gundersen2021semiconditional,ray2024conditionalgan,padmanabha2021conditional,karumuri2024amortized}. Specifically, diffusion generates samples by progressively removing noise. Flow matching instead learns how samples move from a simple source to the target distribution and generates them by integrating an ODE \cite{song2021scorebased,lipman2023flow,tong2024improving}. Their well-defined dynamics and ability to represent high-dimensional distributions have made both approaches increasingly prominent in inverse problems \cite{baldassari2023conditional,chen2025conditionaldiffusions,dasgupta2025conditionalscore,wildberger2023flowmatching,zhai2026conditionalflow}.

Both families can instead combine an unconditional prior with measurements through inference-time guidance. Diffusion guidance modifies the reverse score or drift. DPS estimates a measurement-based correction from a denoised prediction \cite{chung2023dps}. Flow guidance corrects the ODE velocity or optimizes the initial noise \cite{benhamu2024dflow}. These mechanisms are related. Because the observation model is introduced after training, the same prior can often serve different inverse problems when the required likelihood or physics information is available. Guided sampling has been applied to sparse dynamics, turbulent flows, PDE solutions, and structural response fields \cite{gao2024bayesianconditional,li2024s3gm,du2024confield,huang2024diffusionpde,jiang2025odedps,feng2026fullrange}. Approximate corrections may nevertheless miss the intended conditional posterior \cite{feng2025guidance,boys2024tweedie,zach2026dpscalibration}. MoRF therefore follows direct conditional training but reduces the learning burden by fitting only the conditional residual velocity around an analytic Gaussian reference.

\subsection{Operational shift and posterior-spread adaptation}
\label{sec:related-operational-shift}

Operational adaptation methods differ in the target-domain information they use after deployment. Supervised transfer updates predictors using sparse or periodic labels for selected target responses \cite{kadlec2011adaptation,liu2019domainsoftsensor,zhang2023dgman}. Unsupervised methods use unlabeled target sensor records to align inputs or representations \cite{gardner2020domainadaptation,farahani2021dannr,zhang2022damgp,shi2022sensoruda}. These studies demonstrate adaptation without target response labels, with objectives centered mainly on point prediction or representation robustness. Full-field target responses would permit direct updating of field predictors or conditional distributions. They are unavailable in the present setting, where histories from the installed sensors are the only additional target-domain information for adaptation.

When reconstruction is probabilistic, adaptation must also address posterior spread and coverage. Existing work corrects amortized latent posteriors from individual out-of-distribution observations \cite{siahkoohi2023reliable} or calibrates prediction sets under covariate shift \cite{qiu2023predictionsets,yang2024doublyrobust}. For Gaussian measures, Bures-Wasserstein geometry provides an established covariance metric and its associated quadratic-cost transport \cite{gelbrich1990wasserstein,bhatia2019bures}. Related adaptation methods use optimal-transport or Bures objectives to align sample, feature, or class-conditional distributions \cite{courty2017optimaltransport,ren2023buresnet,wang2024jointwasserstein}. Unlike supervised full-field retraining or representation alignment, AST uses installed-sensor histories as its only additional target-domain information to adapt scale-dominated posterior spread while preserving the base-posterior mean and source-trained generator.
 \section{Methodology}
\label{sec:methodology}

\subsection{Engineering Problem and Method Overview}
\label{sec:methodology-overview}

We define probabilistic virtual sensing as the reconstruction of the conditional full-field response distribution \(p(X\mid y)\) from a limited number of noisy sensor measurements. For each event, \(X\in\mathbb R^P\) denotes the multichannel physical response field with \(P\) field entries, and \(y\in\mathbb R^p\) contains the \(p\) scalar sensor readings available for reconstruction. To place response channels with different units and magnitudes on a common scale, the invertible operator \(\mathcal S:\mathbb R^P\rightarrow\mathbb R^P\) divides each field entry by the source-fixed scale of its channel. This gives the dimensionless field \(z=\mathcal S(X)\), while \(\mathcal S^{-1}\) restores physical units. The installed sensors observe selected components of \(z\),
\begin{equation}
    y=Hz+\epsilon,
    \qquad
    \epsilon\sim\mathcal N(0,R),
    \label{eq:full-field-task}
\end{equation}
where the binary matrix \(H\in\mathbb R^{p\times P}\) selects the installed sensor channels, \(\epsilon\in\mathbb R^p\) is the measurement error, and \(R\in\mathbb R^{p\times p}\) is its positive-definite covariance. Because \(y\) contains only sparse information about \(z\), reconstruction is performed in a retained \(d\)-dimensional proper orthogonal decomposition (POD) subspace. The encoder \(\mathcal E:\mathbb R^P\rightarrow\mathbb R^d\) centers and projects \(z\) to the normalized modal coefficient \(c=\mathcal E(z)\), where \(d\) is the number of retained modes. The affine decoder \(\mathcal D:\mathbb R^d\rightarrow\mathbb R^P\) maps a modal coefficient back to a physical response field, including the inverse scaling \(\mathcal S^{-1}\). Given \(y\), the reconstruction method generates posterior coefficient samples \(c^{(n)}\) and decodes them as \(X^{(n)}=\mathcal D(c^{(n)})\), where \(n\) indexes the samples. Their collection represents \(p(X\mid y)\) within the retained POD subspace. In deployment, an operating-condition shift can miscalibrate the posterior spread even when the reconstructed field remains accurate. Without full-field reference data, this spread must be calibrated from the historical sensor context \(\mathcal C\), comprising measurements collected by the same sensors under the current operating condition.

\begin{figure}[htbp]
    \centering
    \includegraphics[width=\textwidth]{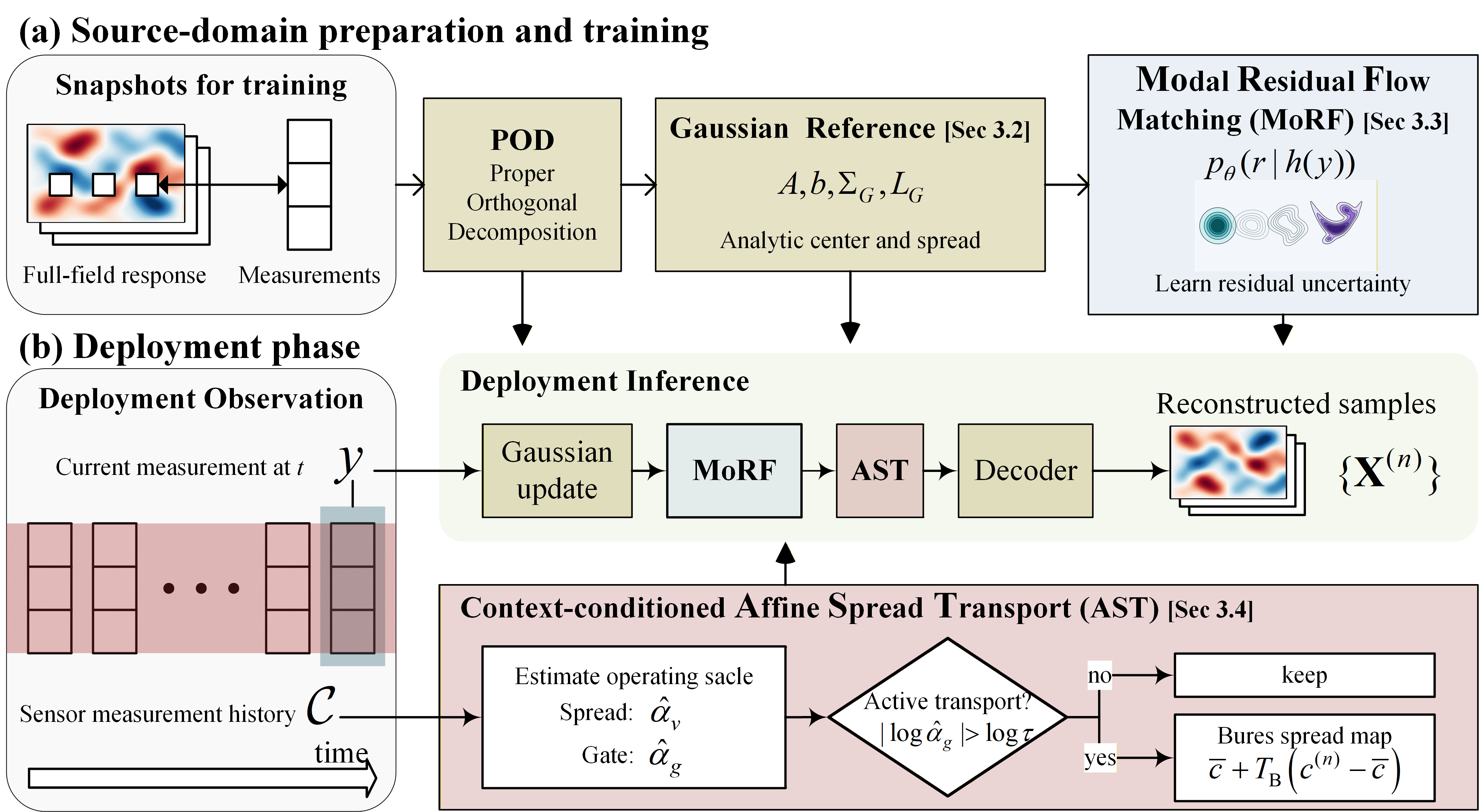}
    \caption{Modal residual flow matching with context-conditioned affine spread transport}
    \label{fig:morf-ast-overview}
\end{figure}

We propose MoRF-AST, a probabilistic virtual sensing framework that combines Modal Residual Flow Matching (MoRF) with Context-Conditioned Affine Spread Transport (AST). As shown in Fig.~\ref{fig:morf-ast-overview}, source full-field data establish the channel scaling, POD representation, Gaussian reference, and decoder (Section~\ref{sec:gaussian-reference}). They also train MoRF (Section~\ref{sec:morf}) to learn the conditional residual distribution beyond the Gaussian reference. Given a current measurement \(y\), the Gaussian update and MoRF produce a base posterior in modal space. AST (Section~\ref{sec:ast}) uses \(\mathcal C\) to assess whether the response scale has changed. It retains the base posterior when the scale remains consistent with the source domain and otherwise adjusts its spread through a symmetric Bures map. The final modal samples are decoded into physical response fields.

 \subsection{Gaussian Reference Model}
\label{sec:gaussian-reference}

The dimensionless source snapshots \(z=\mathcal S(X)\) are centered and arranged as columns of a snapshot matrix. Their proper orthogonal decomposition (POD) is obtained by singular value decomposition \cite{sirovich1987coherent}. Let \(\bar z\in\mathbb R^P\) be the source mean, let \(V\in\mathbb R^{P\times d}\) contain the retained orthonormal left singular vectors, and let \(\Lambda=\operatorname{diag}(\lambda_1,\ldots,\lambda_d)\succ0\) contain the corresponding modal variances. The symbol \(I_d\) denotes the \(d\times d\) identity matrix. The normalized modal encoder, scaled POD loading matrix \(\Phi\in\mathbb R^{P\times d}\), retained field, and physical decoder are
\begin{equation}
    \begin{aligned}
        c=\mathcal E(z)
        &=
        \Lambda^{-1/2}V^\top(z-\bar z),
        &
        \Phi
        &=
        V\Lambda^{1/2},
        \\
        \widehat z(c)
        &=
        \bar z+\Phi c,
        &
        \mathcal D(c)
        &=
        \mathcal S^{-1}\!\left(\widehat z(c)\right).
    \end{aligned}
    \label{eq:pod-decoder}
\end{equation}
The source coefficients have zero empirical mean and identity covariance under this normalization. The affine decoder reconstructs the retained dimensionless field and then restores physical units through \(\mathcal S^{-1}\).

Applying the installed sensor selection to Eq.~\eqref{eq:pod-decoder} gives the retained observation model
\begin{equation}
    A=H\Phi\in\mathbb R^{p\times d},
    \qquad
    b=H\bar z\in\mathbb R^p,
    \qquad
    y=b+Ac+\epsilon.
    \label{eq:pod-observation-model}
\end{equation}
Here \(A\) is the sensor-to-mode observation matrix and \(b\) is the source mean restricted to the installed sensors. Each column of \(A\) gives the sensor response of one normalized mode and therefore records how strongly that structural response pattern is observed.

For source event \(i\), \(z_i\) and \(c_i=\mathcal E(z_i)\) denote the scaled field and modal coefficient, and \(y_i^{\mathrm{clean}}=Hz_i\) is the noise-free sensor reading. Its sensor-space POD truncation residual \(e_i\) satisfies
\[
    y_i^{\mathrm{clean}}
    =
    b+Ac_i+e_i,
    \qquad
    e_i
    =
    H\!\left[z_i-\widehat z(c_i)\right].
\]
The Gaussian reference uses the retained observation model \(y=b+Ac+\epsilon\). The truncation residual has no unique coefficient-space representation within the \(d\) retained modes. Section~\ref{sec:ast} therefore uses its source covariance only when deciding whether the operating context departs from the source reference.

A linear Gaussian model defines the analytic reference for this underdetermined reconstruction,
\begin{equation}
    c\sim\mathcal N(0,I_d),
    \qquad
    \epsilon\sim\mathcal N(0,R).
    \label{eq:gaussian-reference-model}
\end{equation}
POD normalization gives the empirical second-order structure used by the prior, while the standard-normal joint law completes the Gaussian assumption. Using subscript \(G\) for this reference, the conditional law is exactly
\[
    p_G(c\mid y)
    =
    \mathcal N\!\left(\mu_G(y),\Sigma_G\right).
\]
Its covariance \(\Sigma_G\in\mathbb R^{d\times d}\), lower Cholesky factor \(L_G\in\mathbb R^{d\times d}\), and mean \(\mu_G(y)\in\mathbb R^d\) are
\begin{equation}
    \begin{aligned}
        \Sigma_G
        &=
        \left(I_d+A^\top R^{-1}A\right)^{-1}
        =
        L_GL_G^\top,
        \\
        \mu_G(y)
        &=
        \Sigma_GA^\top R^{-1}(y-b).
    \end{aligned}
    \label{eq:gaussian-reference-covariance}
\end{equation}
For a fixed sensor arrangement and noise model, \(A\) and \(R\) are fixed. The covariance \(\Sigma_G\) and its factor \(L_G\) are therefore measurement-independent and can be computed once before deployment. Only the linear update \(\mu_G(y)\) changes with the current measurement. This separation avoids a measurement-specific covariance inversion and gives MoRF a common posterior coordinate system across all source and operating events.

The departure from the analytic center is expressed in the whitened residual coordinate \(r\in\mathbb R^d\),
\begin{equation}
    r=L_G^{-1}\left[c-\mu_G(y)\right].
    \label{eq:posterior-whitened-residual}
\end{equation}
This coordinate measures how far the modal state lies from the analytic estimate, with magnitude and modal correlation normalized by the reference posterior scale. Under the linear Gaussian reference, \(r\mid y\) follows \(\mathcal N(0,I_d)\) exactly. MoRF is trained on the observed source departures from this reference, as described next.
 \subsection{Modal Residual Flow Matching}
\label{sec:morf}

\subsubsection{Conditional Flow Matching}

Flow matching parameterizes a transport from independent Gaussian draws to residual samples conditioned on the current sensor measurement \cite{lipman2023flow,tong2024improving}. Let \(h\in\mathbb R^d\) denote the conditioning vector defined in Eq.~\eqref{eq:morf-condition}. A residual endpoint \(r_1\in\mathbb R^d\) follows the source conditional distribution \(p_{\mathrm{data}}(r\mid h)\), while \(r_0\in\mathbb R^d\) is an independent standard Gaussian start. The interpolation coordinate \(t\in[0,1]\) defines the straight path
\begin{equation}
    \begin{aligned}
        r_0&\sim\mathcal N(0,I_d),
        &
        r_1&\sim p_{\mathrm{data}}(r\mid h),
        \\
        t&\sim\mathcal U(0,1),
        &
        r_t&=(1-t)r_0+tr_1.
    \end{aligned}
    \label{eq:cfm-path}
\end{equation}
The velocity of this path is the constant endpoint displacement \(r_1-r_0\). The neural velocity \(v_\theta\), with trainable parameters \(\theta\), predicts this displacement from the path location, interpolation coordinate, and condition. Its flow-matching loss is
\begin{equation}
    \mathcal L_{\mathrm{FM}}(\theta)
    =
    \mathbb E_{\substack{
        (r_1,h)\sim p_{\mathrm{data}}\\
        r_0\sim\mathcal N(0,I_d),\ t\sim\mathcal U(0,1)
    }}
    \left[
        \frac{1}{d}
        \left\|
        v_\theta(t,r_t,h)-(r_1-r_0)
        \right\|_2^2
    \right].
    \label{eq:cfm-loss}
\end{equation}
Minimizing Eq.~\eqref{eq:cfm-loss} trains \(v_\theta\) to approximate the condition-dependent path velocity. At sampling, the resulting ordinary differential equation
\[
    \frac{\mathrm d r}{\mathrm d t}=v_\theta(t,r,h)
\]
is integrated from \(t=0\) to \(t=1\) to generate residual samples from the learned transport.

\subsubsection{Source Training}

MoRF conditions the residual flow on the standardized analytic posterior mean
\begin{equation}
    h(y)
    =
    \left[\mu_G(y)-\bar\mu\right]\oslash\sigma
    \in\mathbb R^d,
    \label{eq:morf-condition}
\end{equation}
where \(\oslash\) denotes componentwise division. The vector \(\bar\mu\) is the componentwise training mean of \(\mu_G\), and the positive vector \(\sigma\) contains the corresponding standard deviations. Both are computed once from a fixed noise-augmented version of the source training observations. For fixed \(A\) and \(R\), the Gaussian conditional law depends on the measurement only through \(\mu_G(y)\). Thus \(h(y)\) retains this information while providing source-standardized values to the flow.

Each source event provides a clean sensor response \(y_i^{\mathrm{clean}}\) and a modal coefficient \(c_i\). During training, fresh measurement noise is drawn for every batch, and the resulting noisy reading determines both the analytic center and the condition. Equation~\eqref{eq:posterior-whitened-residual} then gives the corresponding residual endpoint, which enters Eq.~\eqref{eq:cfm-loss} to update \(\theta\). This construction aligns training with the noisy measurements available after deployment. MoRF parameterizes \(p_\theta(r\mid h(y))\) to capture the residual bias, modal dependence, and non-Gaussian structure beyond the Gaussian reference. If the reference were exact, the residuals would already follow the standard-normal law and no additional correction would be needed.

\subsubsection{Posterior Sampling}

The current measurement \(y\) enters inference with acquisition noise already present. The frozen velocity transports standard Gaussian points according to the learned residual flow. Multiplication by the Cholesky factor then returns these samples to modal coordinates around the analytic center,
\[
    c^{(n)}
    =
    \mu_G(y)+L_Gr^{(n)}.
\]
Here \(r^{(n)}\) is the terminal residual of trajectory \(n\) at \(t=1\). Numerical integration with \(N_t\) steps produces this endpoint. These samples form the MoRF base posterior represented by \(p_\theta(r\mid h(y))\). The identity branch decodes \(X^{(n)}=\mathcal D(c^{(n)})\), while active AST modifies coefficient spread before decoding.
 \subsection{Context-Conditioned Affine Spread Transport}
\label{sec:ast}

AST uses historical measurements collected at the same sensor locations under the current operating condition to determine whether the MoRF posterior spread should be retained, contracted, or expanded. Its context is \(\mathcal C=\{y_j\}_{j=1}^{N_c}\), where \(N_c\ge1\) is the number of past sensor measurements and each \(y_j\in\mathbb R^p\) contains acquisition noise. From this context, AST estimates an operating scale shared by current measurements, applies an identity gate, and maps the deviations of the base coefficient samples when a scale change is detected. It requires no full-field data from the current operating condition and leaves the posterior mean unchanged.

\subsubsection{Effective Operating Scale Estimation}

AST represents an operating change by scaling the normalized source modal prior while retaining the source mean and POD basis,
\[
    c\sim\mathcal N(0,\alpha I_d),
    \qquad
    \alpha>0.
\]
Under the installed sensor arrangement, the corresponding source signal covariance is
\[
    S_y=AA^\top\in\mathbb R^{p\times p},
\]
and its operating-scale counterpart is \(\alpha S_y\). The scalar \(\alpha\) therefore describes an overall change in response amplitude relative to the source reference.

Let \(\mathcal G\subset\mathbb R_{>0}\) be a logarithmic grid of candidate scales, and define the centered context measurement \(\delta y_j=y_j-b\). For a positive-semidefinite sensor-space signal covariance \(\Gamma\in\mathbb R^{p\times p}\), the mean-fixed Gaussian negative log-likelihood and its scale estimate are
\begin{equation}
    \begin{aligned}
        J(\alpha,\Gamma)
        &=
        \frac{1}{2N_c}
        \sum_{j=1}^{N_c}
        \left[
            \delta y_j^\top
            (R+\alpha\Gamma)^{-1}\delta y_j
            +
            \log\det(R+\alpha\Gamma)
        \right],
        \\
        \widehat\alpha(\Gamma)
        &=
        \underset{\alpha\in\mathcal G}{\arg\min}\,
        J(\alpha,\Gamma).
    \end{aligned}
    \label{eq:ast-scale-objective}
\end{equation}
The estimate is shared by measurements from the same operating condition. Because the context is centered about the source sensor mean \(b\),
\[
    \mathbb E[\delta y\delta y^\top]
    =
    \operatorname{Cov}(y)
    +
    (\mathbb E[y]-b)(\mathbb E[y]-b)^\top.
\]
The fitted \(\widehat\alpha\) thus captures both covariance change and the contribution of an operating mean offset. It is an effective response-amplitude scale rather than a pure covariance ratio.

\subsubsection{Truncation-Aware Identity Decision}

The source POD truncation residuals \(e_i\) defined in Section~\ref{sec:gaussian-reference} quantify the response variance omitted from the retained modes at the sensor locations. From \(N_{\mathrm{cal}}\) source calibration records, their covariance is estimated once as
\[
    \bar e
    =
    \frac{1}{N_{\mathrm{cal}}}
    \sum_{i=1}^{N_{\mathrm{cal}}}e_i,
    \qquad
    \Sigma_{\mathrm{tr}}
    =
    \frac{1}{N_{\mathrm{cal}}}
    \sum_{i=1}^{N_{\mathrm{cal}}}
    (e_i-\bar e)(e_i-\bar e)^\top.
\]
Here \(\Sigma_{\mathrm{tr}}\in\mathbb R^{p\times p}\) is the sensor-space covariance of the POD truncation residual, while \(R\) is the measurement-error covariance.

AST evaluates the scale estimator in two forms. The gate estimate \(\widehat\alpha_g\) includes \(\Sigma_{\mathrm{tr}}\) when deciding whether transport is needed, while the transport estimate \(\widehat\alpha_v\) uses only the covariance represented by the retained modes. With a tolerance \(\tau>1\), these estimates and the binary gate \(g\in\{0,1\}\) are
\begin{equation}
    \begin{aligned}
        \widehat\alpha_v
        &=
        \widehat\alpha(S_y),
        &
        \widehat\alpha_g
        &=
        \widehat\alpha(S_y+\Sigma_{\mathrm{tr}}),
        \\
        g
        &=
        \mathbf 1
        \left(
            \left|\log\widehat\alpha_g\right|>\log\tau
        \right).
    \end{aligned}
    \label{eq:ast-gate}
\end{equation}
The interval \([\tau^{-1},\tau]\) defines an identity band. Values within the band retain the MoRF base posterior, while values outside it activate transport. Including \(\Sigma_{\mathrm{tr}}\) in the gate prevents omitted source variability from being mistaken for an operating change. It is excluded from the transport because a general sensor-space truncation covariance has no unique representation within the retained coefficient space.

\subsubsection{Mean-Preserving Affine Spread Transport}

When transport is active, \(\widehat\alpha_v\) defines the scaled-reference posterior covariance
\begin{equation}
    \Sigma_v
    =
    \left(
        \widehat\alpha_v^{-1}I_d
        +
        A^\top R^{-1}A
    \right)^{-1}.
    \label{eq:ast-target-covariance}
\end{equation}
The symmetric Bures map \cite{gelbrich1990wasserstein} from the source reference covariance \(\Sigma_G\) to \(\Sigma_v\) is
\begin{equation}
    T_{\mathrm B}
    =
    \Sigma_G^{-1/2}
    \left(
        \Sigma_G^{1/2}
        \Sigma_v
        \Sigma_G^{1/2}
    \right)^{1/2}
    \Sigma_G^{-1/2},
    \label{eq:ast-bures-map}
\end{equation}
where every symmetric matrix square root is the principal square root. For positive-definite \(\Sigma_G\) and \(\Sigma_v\), this is the unique symmetric positive-definite map satisfying \(T_{\mathrm B}\Sigma_GT_{\mathrm B}^\top=\Sigma_v\). It is also the quadratic-cost \(W_2\)-optimal transport between same-mean Gaussian distributions with these covariances.

Although \(\widehat\alpha_v\) is a scalar, conditioning makes the correction direction dependent. Writing
\[
    A^\top R^{-1}A
    =
    \Psi\operatorname{diag}(\rho_1,\ldots,\rho_d)\Psi^\top,
\]
the Bures variance multiplier along direction \(i\) is
\[
    t_i^2
    =
    \frac{
        \widehat\alpha_v(1+\rho_i)
    }{
        1+\widehat\alpha_v\rho_i
    }.
\]
The sign of the variance change is determined by \(\widehat\alpha_v-1\), while its magnitude decreases as the sensor information \(\rho_i\) increases. AST therefore acts most strongly along weakly observed directions and is attenuated where the measurements already constrain the response.

For a current measurement, let \(N_s\) be the number of MoRF base samples \(c^{(n)}\). Their empirical mean and the final modal samples are
\begin{equation}
    \begin{aligned}
        \bar c
        &=
        \frac{1}{N_s}
        \sum_{n=1}^{N_s}
        c^{(n)},
        \\
        \widetilde c^{(n)}
        &=
        \begin{cases}
            \bar c+
            T_{\mathrm B}
            \left(c^{(n)}-\bar c\right),
            & g=1,
            \\[2mm]
            c^{(n)},
            & g=0.
        \end{cases}
    \end{aligned}
    \label{eq:ast-sample-map}
\end{equation}
Because the map acts on deviations about \(\bar c\), it preserves the empirical coefficient mean and, through the affine decoder \(\mathcal D\), the posterior mean field. The final samples are \(X^{(n)}=\mathcal D(\widetilde c^{(n)})\).

The construction replaces the Gaussian reference covariance while retaining the covariance structure learned by MoRF relative to that reference. Since \(\Sigma_G\) and \(\Sigma_v\) are both matrix functions of \(A^\top R^{-1}A\), they commute. For any empirical base covariance \(C\),
\begin{equation}
    \Sigma_v^{-1/2}
    T_{\mathrm B}CT_{\mathrm B}^\top
    \Sigma_v^{-1/2}
    =
    \Sigma_G^{-1/2}C\Sigma_G^{-1/2}.
    \label{eq:ast-normalized-covariance}
\end{equation}
AST therefore changes the reference spread without discarding the reference-normalized covariance structure of the base posterior. Section~\ref{sec:domain-calibration} evaluates how this structure interacts with the transport direction and the base spread bias. The complete procedure is given in Algorithm~\ref{alg:morf-ast-reconstruction}.

\begin{algorithm}[H]
    \caption{MoRF-AST probabilistic full-field reconstruction}
    \label{alg:morf-ast-reconstruction}
    \begin{algorithmic}[1]
        \Require Current measurement \(y\) and historical context \(\mathcal C\)
        \Require Trained MoRF model, \(N_s\) samples, and \(N_t\) ODE steps
        \State \(\widehat\alpha_v\gets\widehat\alpha(S_y)\), \(\widehat\alpha_g\gets\widehat\alpha(S_y+\Sigma_{\mathrm{tr}})\) \Comment{Scale estimation}
        \State \(r_0^{(n)}\sim\mathcal N(0,I_d)\), \(n=1,\ldots,N_s\) \Comment{Gaussian initialization}
        \State \(r^{(n)}\gets\operatorname{ODESolve}(v_\theta,h(y),r_0^{(n)},N_t)\) \Comment{MoRF sampling}
        \State \(c^{(n)}\gets\mu_G(y)+L_Gr^{(n)}\) \Comment{Base posterior}
        \State \(\bar c\gets N_s^{-1}\sum_{n=1}^{N_s}c^{(n)}\) \Comment{Ensemble center}
        \If{\(\left|\log\widehat\alpha_g\right|>\log\tau\)} \Comment{Transport branch}
            \State \(\Sigma_v\gets(\widehat\alpha_v^{-1}I_d+A^\top R^{-1}A)^{-1}\) \Comment{Target covariance}
            \State \(T_{\mathrm B}\gets\Sigma_G^{-1/2}(\Sigma_G^{1/2}\Sigma_v\Sigma_G^{1/2})^{1/2}\Sigma_G^{-1/2}\) \Comment{Bures map}
            \State \(\widetilde c^{(n)}\gets\bar c+T_{\mathrm B}(c^{(n)}-\bar c)\) \Comment{Active transport}
        \Else \Comment{Identity branch}
            \State \(\widetilde c^{(n)}\gets c^{(n)}\)
        \EndIf
        \Ensure \(\{X^{(n)}=\mathcal D(\widetilde c^{(n)})\}_{n=1}^{N_s}\) \Comment{Field decoding}
    \end{algorithmic}
\end{algorithm}
  \section{Numerical Validation}
\label{sec:numerical-validation}

\subsection{Numerical Benchmark}
\label{sec:numerical-benchmark}

\subsubsection{Bridge Deck and Structural Model}
\label{sec:bridge-response-model}

\begin{figure}[htbp]
    \centering
    \includegraphics[width=\textwidth]{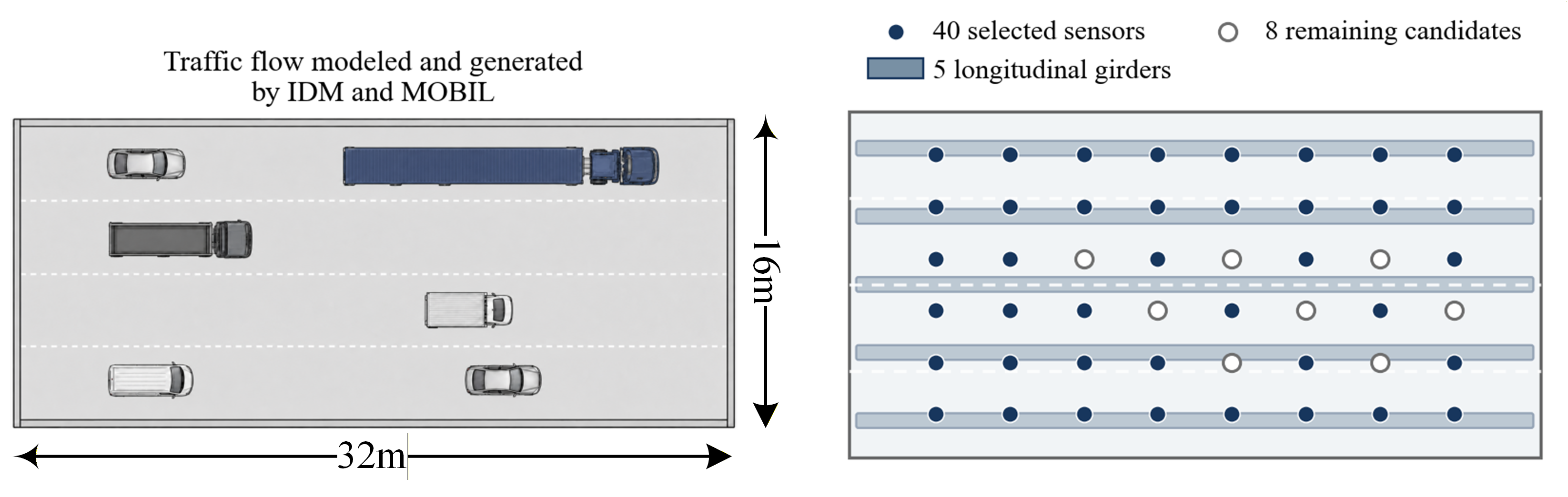}
    \caption{Bridge benchmark for probabilistic full-field reconstruction}
    \label{fig:numerical-benchmark}
\end{figure}

We consider probabilistic full-field reconstruction of a bridge deck from sparse multichannel observations. Figure~\ref{fig:numerical-benchmark} shows the \(32~\mathrm{m}\times16~\mathrm{m}\), four-lane deck with five distributed longitudinal girders. On a \(128\times128\) grid, its deflection satisfies \cite{timoshenko1959plates}
\begin{equation}
    \left[D_x+\kappa_g(y)\right]w_{xxxx}
    +2\left(D_1+2D_k\right)w_{xxyy}
    +D_y w_{yyyy}
    =
    q(x,y).
    \label{eq:bridge-plate}
\end{equation}
Here \(x\) and \(y\) denote the longitudinal and transverse coordinates, \(w\) is the deflection, and \(q\) is the wheel-load pressure. The coefficients \(D_x\) and \(D_y\) are the flexural rigidities, \(D_1\) and \(D_k\) are the coupling and twisting rigidities, and \(\kappa_g\) represents the girder contribution. Subscripts denote partial derivatives. The deck has simply supported ends and free side edges. Traffic realizations are generated by a seeded four-lane microscopic simulator with regime-specific distributions for lane flow, heavy-vehicle composition, speed, headway, and lateral position. Vehicle weights are allocated to axles and wheel pairs, and the on-deck wheel forces are rasterized as normalized Gaussian pressure patches to form \(q(x,y)\), with each pressure field defining one event.

Each event comprises seven response fields in the fixed order
\begin{equation}
    \left(
    w,\kappa_x,\kappa_y,\kappa_{xy},M_x,M_y,M_{xy}
    \right).
    \label{eq:response-channels}
\end{equation}
The model reconstructs all seven fields, but the quantitative analysis focuses on \(M_x\), \(M_y\), and \(M_{xy}\). The deflection field is comparatively easy to reconstruct, and its posterior intervals are already narrow, leaving little scope for AST to improve the spread. The moment fields provide a more demanding test of full-field uncertainty and can be used with the section properties to recover stresses.

Each channel is normalized as \(z_j=X_j/s_j\), where \(s_j\) is the source-training standard deviation pooled over the 40 master locations. The fixed \(d=192\) source POD captures \(99.9\%\) of the variance and is used across all domains.

An \(8\times6\) interior candidate grid is ordered by farthest-point sampling from the deck centre. The first 40 positions form the fixed master sensor layout. Each design with fewer sensors uses the first \(k\) positions in this ordering. The 40-location layout provides space-filling coverage while preserving the intended underdetermination.

The operational domains in Table~\ref{tab:operational-domains} isolate changes in traffic loading statistics. The bridge model, boundary conditions, response normalization, POD representation, and observation design remain fixed. The source data \(S\) and E0 are independent draws from the same weighted mixture of nine traffic regimes. The models therefore learn from a broad source prior, while E0 provides an independent control over the full source envelope. N0, N1, and V1 to V6 draw from individual regime components to represent operating periods dominated by a particular traffic pattern.

\begin{table}[htbp]
    \centering
    \caption{Operational domains and traffic regimes}
    \label{tab:operational-domains}
    \footnotesize
    \setlength{\tabcolsep}{4pt}
    \begin{tabularx}{\textwidth}{@{}l
        >{\raggedright\arraybackslash}p{0.22\textwidth}
        >{\raggedright\arraybackslash}X@{}}
        \toprule
        Domain & Dominant regime & Operating meaning \\
        \midrule
        E0 & Source mixture & Independent source-envelope control \\
        N0 & Restricted load & Heavy vehicles largely excluded \\
        N1 & Night & Low total flow with a non-negligible heavy-vehicle share \\
        V1 & Commuter peak & Dense, slow commuter traffic \\
        V2 & Platoon & Clustered convoy arrivals \\
        V3 & Work zone & Concentrated lane use with an elevated heavy-vehicle share \\
        V4 & Freight & High heavy-vehicle share \\
        V5 & Overload & Traffic concentrated toward the heaviest vehicle classes \\
        V6 & Congested & Slow, high-volume freight traffic \\
        \bottomrule
    \end{tabularx}
\end{table}

The source data comprise 8,000 training and 1,000 validation events. Each evaluation domain contains 1,500 sensor-only context events and 800 test events. Context and test identifiers are disjoint at the simulation-run level. Context fields are not retained. Test full-field responses are reserved for offline scoring, held-out diagnostics, and the full-field visualization, and never enter model fitting or AST.

\subsection{Experimental Design}
\label{sec:experimental-design}

\subsubsection{Baselines}
\label{sec:baselines}

We consider three baselines: the analytic Gaussian reference, a Deep Ensemble \cite{lakshminarayanan2017deep}, and Gaussian mixture residual regression (GMR) \cite{calinon2007gmr}. The Gaussian reference in Section~\ref{sec:gaussian-reference} has no fitted network and tests whether learning non-Gaussian residuals improves on the analytic posterior. The Deep Ensemble is a standard supervised uncertainty baseline. Its ten independently trained members take the standardized Gaussian posterior mean as input, predict diagonal Gaussian residual distributions, and contribute equally weighted samples. GMR is a conventional conditional-density baseline. It conditions on the standardized query observation residual and fits a Gaussian mixture to selected high-variance residual coordinates, testing whether a mixture model can reproduce the residual distribution without a flow. This condition is separate from the historical context used by AST. MoRF and MoRF-AST are evaluated against these baselines.

\subsubsection{Experimental Setup}
\label{sec:experimental-setup}

The primary domain comparison changes only the traffic regime and uses the fixed 40-location layout. All methods share the source scaling, POD basis, observation operator, measurement covariance, decoder, evaluation events, and score implementation. Methods expressed in residual coordinates also share the Gaussian reference. Query observations use common noise realizations across methods. Reduced sensor designs use nested prefixes of the same master ordering, so sensor addition is the only change between consecutive layouts.

Target context observations are standardized using the source scales and kept separate from the test events. The source calibration set used to estimate \(\Sigma_{\mathrm{tr}}\) combines the source training and validation fields. The AST identity gate uses \(\tau=1.25\) in Eq.~\eqref{eq:ast-gate}.

The directional coverage analysis pools each coverage cell over the 800 held-out events and 16,344 unobserved grid points. Confidence intervals are estimated from 2,000 paired bootstrap replicates that resample events as clusters and retain all fields, nominal levels, and spatial locations within each event. This construction preserves spatial dependence and pairs the base and adapted scores within every replicate.

MoRF uses fixed-step Heun integration with 100 steps. All methods use 500 posterior samples per observation.

\subsubsection{Metrics}
\label{sec:metrics}

We evaluate posterior accuracy and calibration using the following three metrics. Scores are computed directly in physical coordinates for \(M_x\), \(M_y\), and \(M_{xy}\). Let \(X\) be the true field, \(\widehat X\) and \(\widehat X'\) be independent posterior draws, \(\overline X=\mathbb E[\widehat X]\), and \(X_{\mathrm{rms}}\) be the root-mean-square magnitude of the true field. Angular brackets denote averaging over test events and unobserved grid points. For a nominal level \(a\in\mathcal A=\{0.1,0.2,\ldots,0.9\}\), the empirical coverage \(\widehat C_a\) is the fraction of true values contained in the corresponding central posterior interval. The three metrics are
\begin{align}
    \operatorname{NRMSE}
    &=\frac{\left\langle(\overline X-X)^2\right\rangle^{1/2}}{X_{\mathrm{rms}}},
    \label{eq:metric-nrmse}\\
    \operatorname{nCRPS}
    &=\frac{1}{X_{\mathrm{rms}}}
    \left\langle
    \mathbb E|\widehat X-X|
    -\tfrac{1}{2}\mathbb E|\widehat X-\widehat X'|
    \right\rangle,
    \label{eq:ncrps}\\
    \operatorname{ACE}
    &=\frac{1}{9}\sum_{a\in\mathcal A}
    |\widehat C_a-a|.
    \label{eq:ace}
\end{align}
ACE averages the absolute difference between empirical and nominal coverage over the nine levels. Lower ACE indicates better marginal calibration, with zero denoting exact agreement. To distinguish undercoverage from overcoverage, the directional analysis also uses the signed coverage error
\begin{equation}
    \operatorname{SCE}_{m,d}
    =
    \frac{1}{3|\mathcal A|}
    \sum_{j\in\{M_x,M_y,M_{xy}\}}
    \sum_{a\in\mathcal A}
    \left(
        \widehat C_{m,d,j,a}-a
    \right),
    \label{eq:signed-coverage-error}
\end{equation}
where \(\widehat C_{m,d,j,a}\) is the empirical central-interval coverage for model \(m\), domain \(d\), moment field \(j\), and nominal level \(a\). Negative SCE denotes undercoverage, while positive SCE denotes overcoverage.

The CRPS expectation is evaluated with the fair finite-ensemble correction \cite{ferro2014fair}. Each primary metric is first computed for each of the three moment fields and then averaged across fields. NRMSE and nCRPS are reported with percentage signs, whereas ACE is reported as a fraction. Empirical scores are reported to three significant figures throughout. E0 is reported separately, while averages over N0, N1, and V1 to V6 assign equal weight to each of the eight shifted domains.

\subsection{Calibration Performance across Operational Domains}
\label{sec:domain-calibration}

\begin{figure}[htbp]
    \centering
    \includegraphics[width=\textwidth]{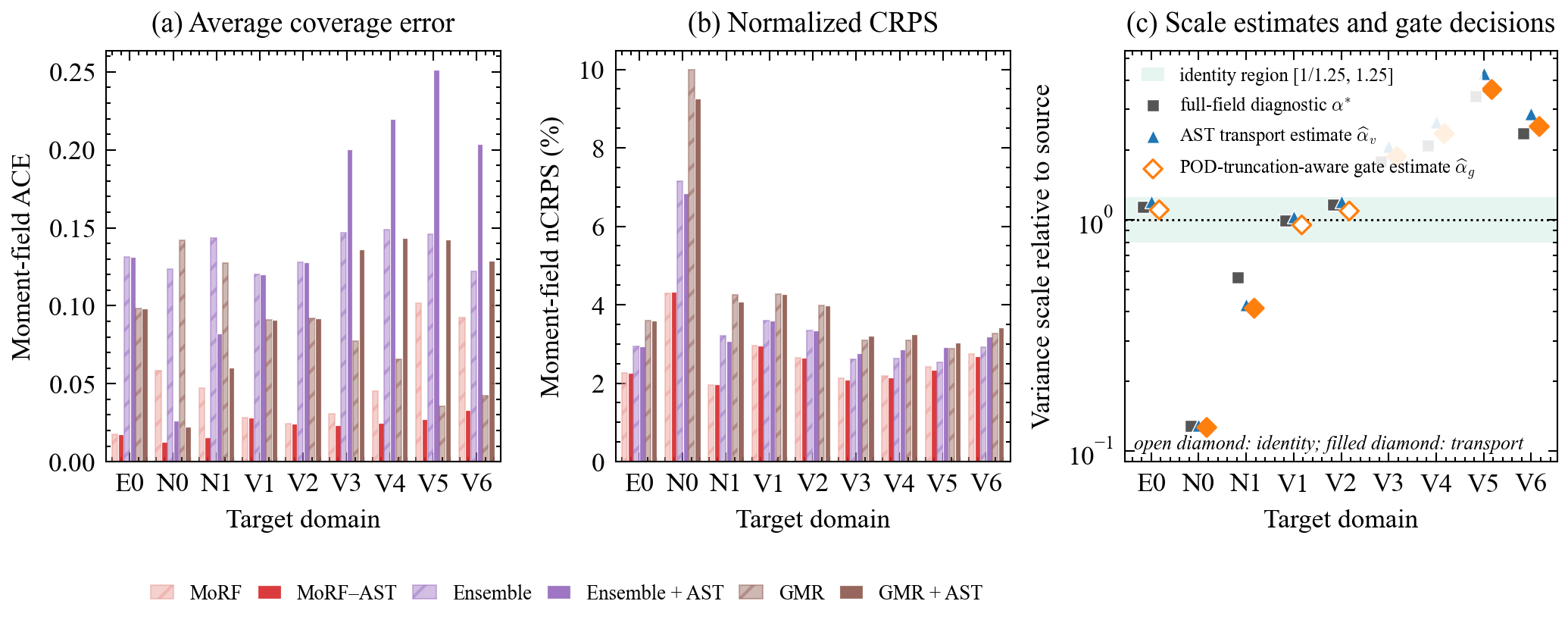}
    \caption{Cross-domain posterior calibration and AST routing}
    \label{fig:domain-calibration}
\end{figure}

We compare MoRF, the Deep Ensemble, and GMR before and after AST across the operational domains using ACE and nCRPS, and record the corresponding scale estimates and gate decisions. The results are summarized in Fig.~\ref{fig:domain-calibration}. The identity gate retains the base posterior in E0, V1, and V2, while Bures transport is activated in N0, N1, and V3 to V6. E0 therefore remains an independent source-envelope control and is reported separately from the shifted-domain average.

Across the eight shifted domains, MoRF-AST reduces the mean ACE from 0.0535 to 0.0236, a relative reduction of \(55.9\%\), and yields the lowest aggregate calibration error in the comparison. The corresponding mean NRMSE remains \(7.72\%\) by construction, while nCRPS changes only from \(2.66\%\) to \(2.64\%\). All six activated domains improve, while the identity-routed domains retain their base values. The primary gain is better marginal coverage, while overall probabilistic accuracy changes only slightly.

The same Bures map does not improve the shifted-domain aggregate ACE of the Deep Ensemble or GMR. The Deep Ensemble mean ACE increases from 0.135 to 0.154, while the GMR value increases from 0.0841 to 0.102. Both alternatives benefit in some domains, but their losses elsewhere dominate the aggregate response. The estimated operating scale is therefore not sufficient to determine whether transport will improve calibration. Its effect also depends on the spread bias already present in the base posterior.

\begin{figure}[htbp]
    \centering
    \includegraphics[width=\textwidth]{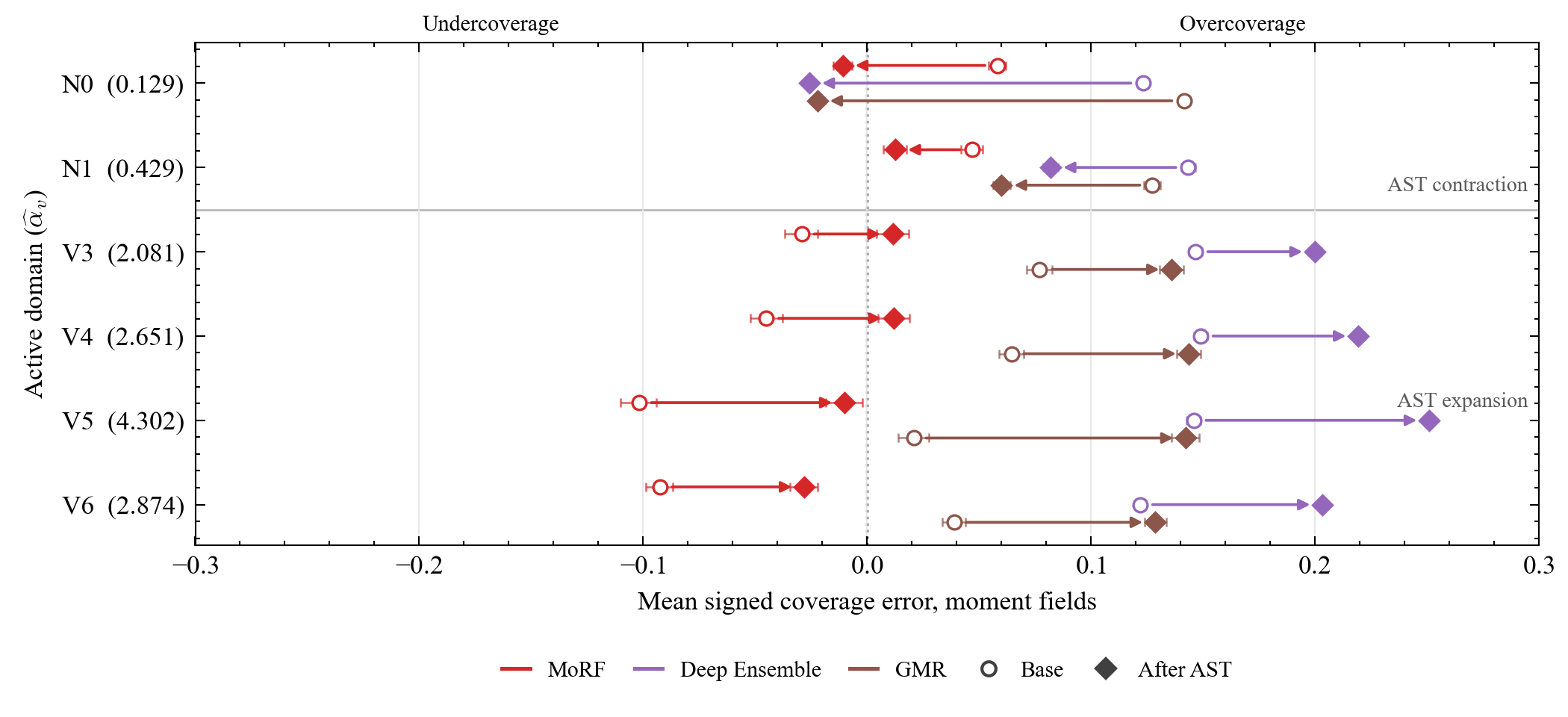}
    \caption{Signed coverage under active Bures transport}
    \label{fig:directional-compatibility}
\end{figure}

Figure~\ref{fig:directional-compatibility} resolves this dependence using the signed coverage error in Eq.~\eqref{eq:signed-coverage-error}. N0 and N1 require reference contraction, whereas V3 to V6 require reference expansion.

\FloatBarrier

In the contraction domains, every base model overcovers and AST moves each signed error toward zero. In the expansion domains, MoRF undercovers and AST again moves the signed error toward zero in all four cases. The Deep Ensemble and GMR already overcover in these domains, so the same expansion moves them farther from nominal coverage. In every activated domain, the estimated paired-bootstrap probability that AST lowers ACE is at least 0.957 for MoRF. For the Deep Ensemble and GMR, the corresponding estimate is 0 in every expansion domain. Contraction is beneficial when the base posterior is too broad, while expansion is beneficial only when it acts on an underdispersed posterior.

This directional compatibility explains the contrasting aggregate results. The Bures map determines contraction or expansion from \(\widehat\alpha_v\) and the sensor geometry, independently of the signed coverage bias of the base model, and carries forward its reference-normalized covariance geometry. MoRF is the only tested model whose base bias is aligned with the required action in both transport regimes. This pattern is consistent with the full-dimensional conditional residual model capturing the principal normalized posterior structure and leaving an operating-scale spread mismatch for AST to correct. The two alternative models enter the expansion domains with an existing positive coverage bias, which the reference expansion amplifies.

\subsection{Full-Field Posterior Visualization}
\label{sec:field-visualization}

\begin{figure}[htbp]
    \centering
    \includegraphics[width=\textwidth]{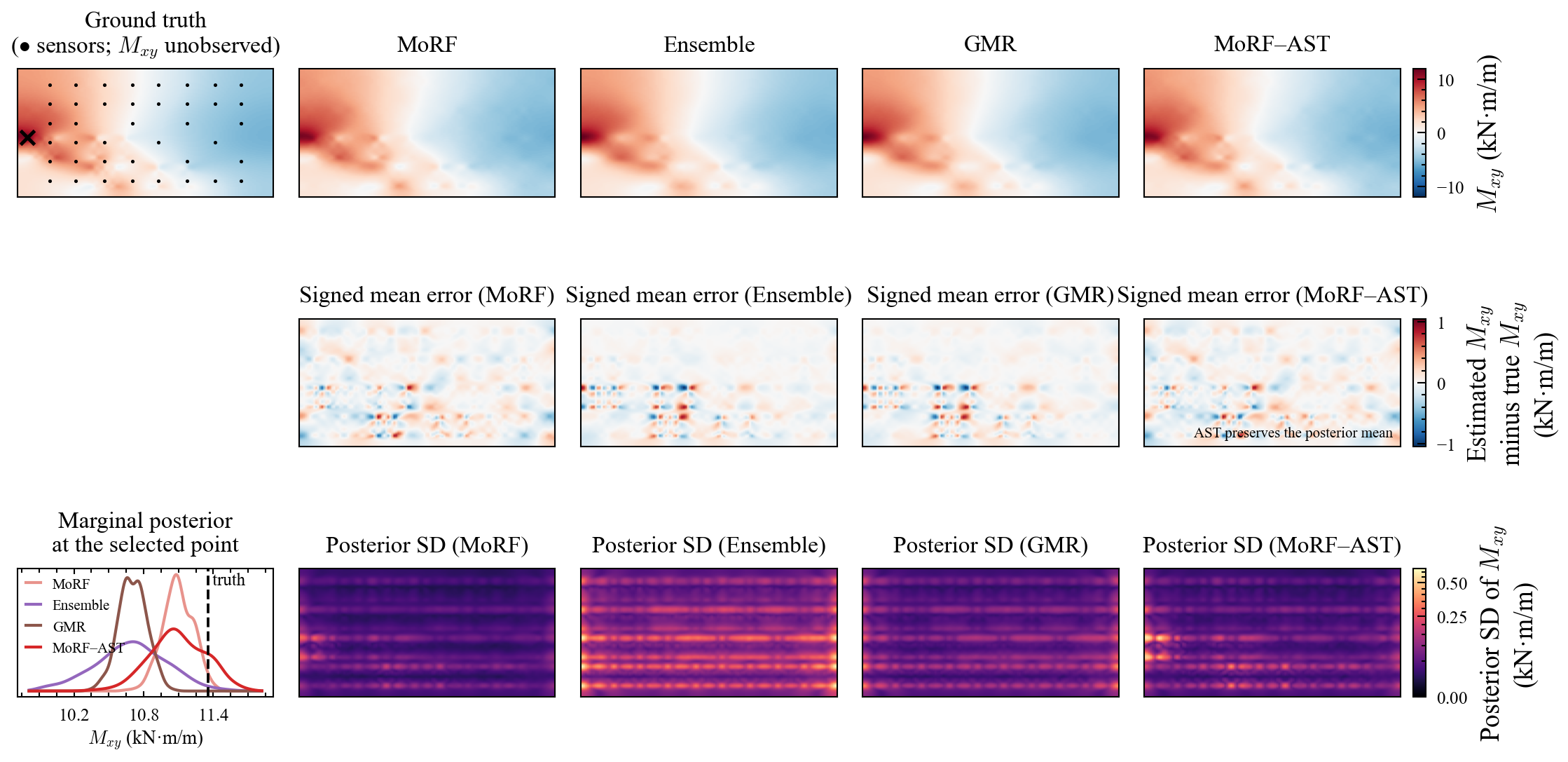}
    \caption{Full-field reconstruction and posterior spread under overload traffic}
    \label{fig:field-showcase}
\end{figure}

To examine the domain-level calibration result in physical space, we select the V5 test event with the largest absolute \(M_{xy}\) response and compare the truth, posterior means, reconstruction errors, peak-location marginals, and posterior standard-deviation fields. The resulting full-field visualization is shown in Fig.~\ref{fig:field-showcase}.

MoRF and MoRF-AST give identical mean and error fields. Both resolve the sign and location of the principal high-response structure. At the peak, the unadapted MoRF 90\% interval ends below the true response. AST expands the distribution about the same posterior center, bringing the truth inside the corresponding interval. The Deep Ensemble distribution is broader than the MoRF-AST distribution, although its center remains below the truth. The GMR distribution is narrower than MoRF-AST, and its center also remains below the truth. Interval width alone is therefore insufficient. The center and spread must be compatible with the event response. The MoRF standard-deviation field concentrates its largest local variation around the high-response region. AST increases the uncertainty scale while retaining this spatial organization. The event visualizes the cross-domain coverage correction in physical space and complements the aggregate test-set evidence.
 \section{Ablation, Sensitivity, and Operational Studies}
\label{sec:ablation-sensitivity-operational}

\subsection{Design Comparisons and Ablation Studies}
\label{sec:design-comparisons}

\subsubsection{Effect of Residual Formulation}
\label{sec:residual-comparison}

To determine whether MoRF's data efficiency arises from Gaussian conditioning alone or from learning a residual around the analytic reference, we compare it with two direct conditional flow models. Direct-\(y\) learns modal coefficients \(c\) from the sparse observation, while direct-\(\mu\) learns \(c\) from the standardized Gaussian posterior mean. MoRF uses the same condition as the parameter-matched direct-\(\mu\) model but learns the posterior-whitened residual \(r\). All three models are trained on the same nested subsets of the 8,000 source events and evaluated on held-out E0 events.

\begin{figure}[htbp]
    \centering
    \includegraphics[width=\textwidth]{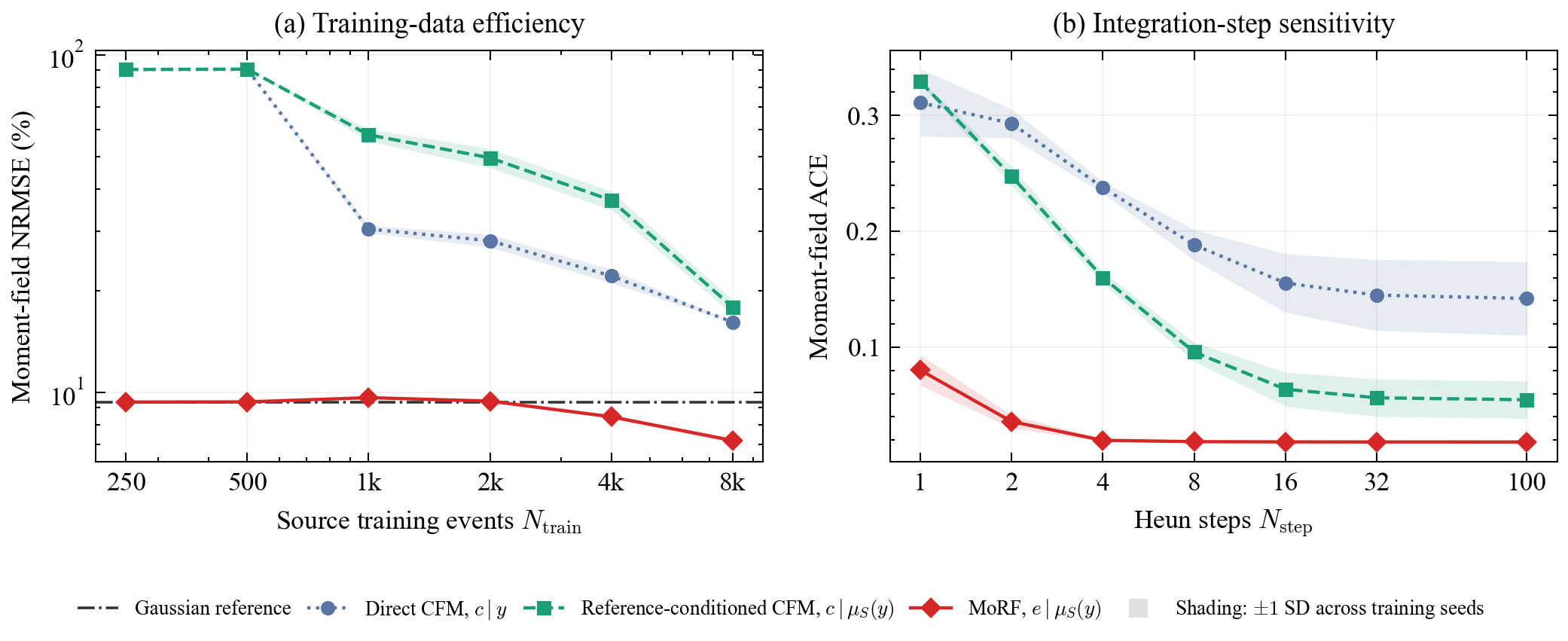}
    \caption{Training-data and integration-step sensitivity on the no-shift control}
    \label{fig:efficiency}
\end{figure}

The comparison across flow-training set sizes in Fig.~\ref{fig:efficiency} separates the contribution of the analytic reference from that of the residual flow. With 250 flow-training events, MoRF obtains \(9.35\%\) posterior-mean NRMSE, nearly identical to the \(9.34\%\) Gaussian reference, while both direct flows remain near \(90.5\%\). The analytic update therefore anchors MoRF when full-field training data are scarce. At 8,000 events, MoRF reaches \(7.20\%\), compared with \(16.1\%\) for direct-\(y\) and \(17.9\%\) for the parameter-matched direct-\(\mu\). The widening performance gap from the fixed reference, rather than a gap from the direct flows, identifies the additional structure learned by the residual model.

The integration step comparison used the full training set, fixed checkpoints, and common base samples. Increasing the integration budget from one to four Heun steps removes most of the integration-induced calibration error, while further refinement to 100 steps changes the result only slightly. Four steps therefore approach the 100-step result with one twenty-fifth as many integration steps. Together, these results show that the analytic reference makes MoRF data-efficient, while the residual flow reaches near-converged calibration with four Heun steps.

\subsubsection{Guided Sampling Comparison}
\label{sec:guidance-comparison}

\begin{figure}[htbp]
    \centering
    \includegraphics[width=\textwidth]{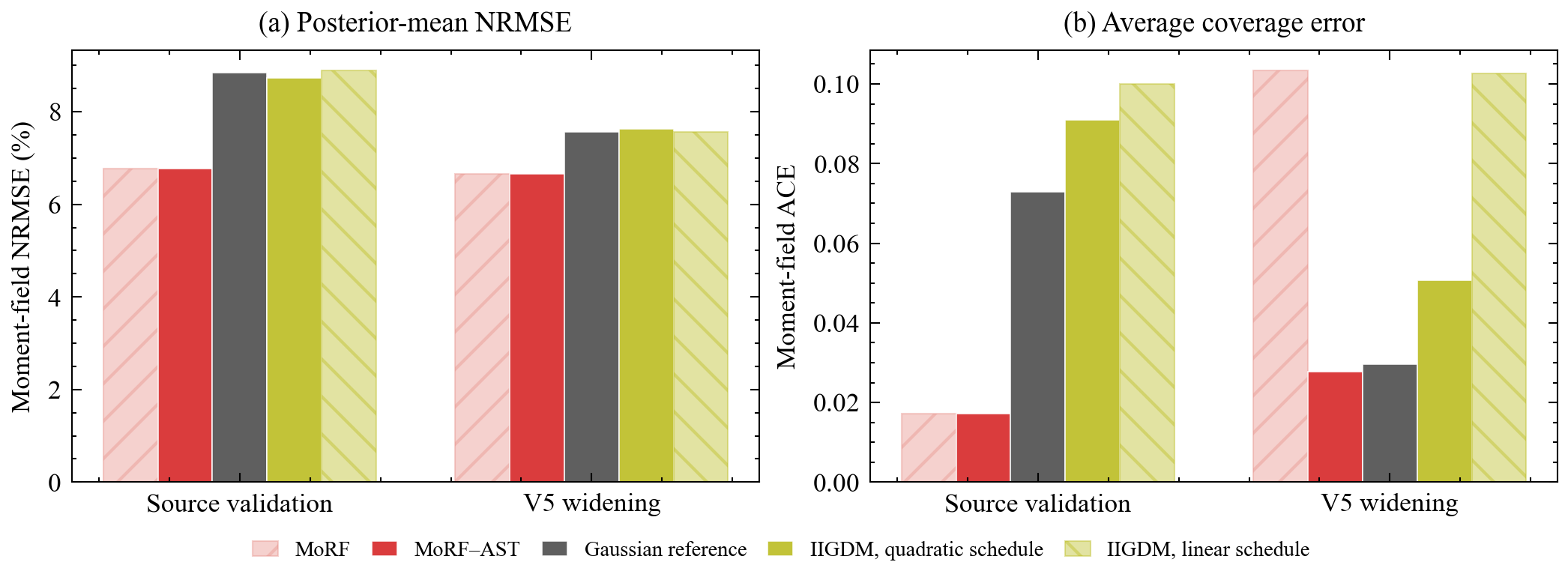}
    \caption{Direct posterior modeling and likelihood guidance across source validation and V5}
    \label{fig:guidance}
\end{figure}

To assess the value of direct conditional posterior learning, we compare MoRF with an unconditional modal flow that incorporates the observation through \(\Pi\)GDM-style likelihood guidance during sampling \cite{song2023pigdm}. The two guidance cases use linearly and quadratically decaying guidance variances with the same 100-step integration budget. We evaluate source validation, where AST returns identity, and V5, where the complete MoRF-AST result includes active spread transport. The results are shown in Fig.~\ref{fig:guidance}.

On source validation, direct conditional MoRF gives the best mean reconstruction and calibration among the displayed flow constructions. In V5, the same conditional model retains its reconstruction advantage, while active AST restores calibration under the operating shift. These results support direct conditional MoRF with AST over likelihood-guided unconditional sampling under the same integration budget.

\subsubsection{Ablation of Posterior Center and Spread Map}
\label{sec:map-ablation}

\begin{figure}[htbp]
    \centering
    \includegraphics[width=\textwidth]{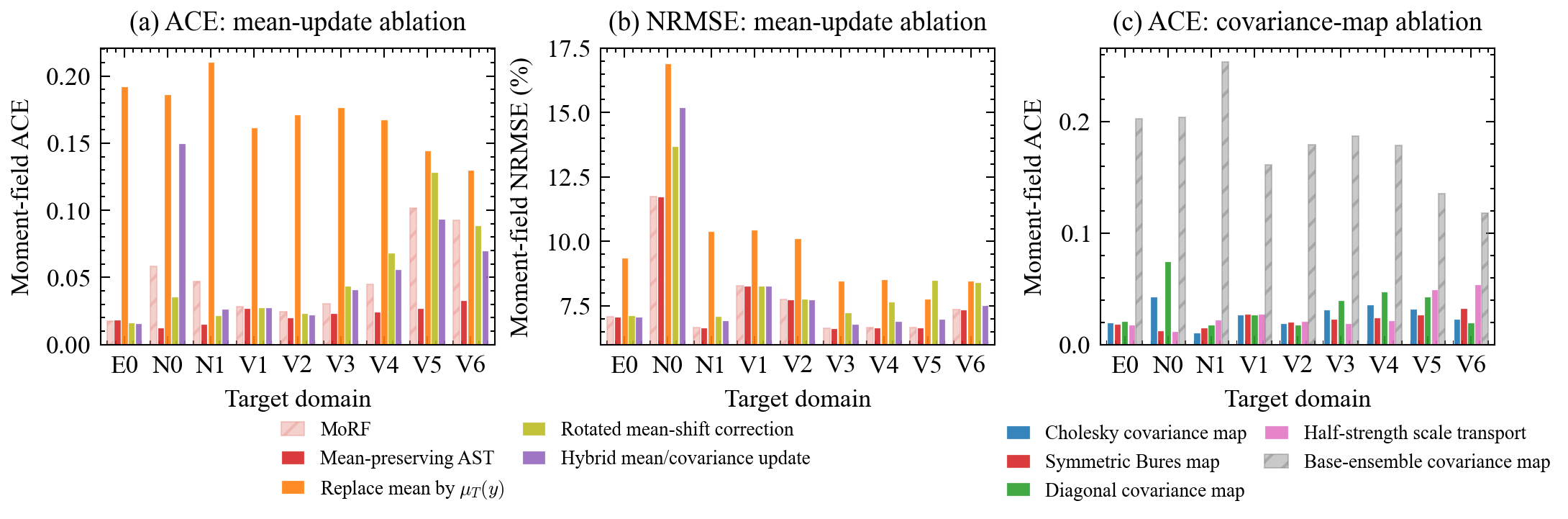}
    \caption{Posterior-center and covariance-map ablations across operational domains}
    \label{fig:map-ablation}
\end{figure}

To isolate the effects of posterior centering and covariance transport, we apply ungated maps to the same MoRF base samples across all nine domains, including E0 (Fig.~\ref{fig:map-ablation}). Every spread-only map leaves NRMSE unchanged within numerical precision. Replacing that center with the target-scale Gaussian mean increases the nine-domain mean NRMSE from \(7.65\%\) to \(10.1\%\) and the mean ACE from 0.0495 to 0.171. Mean preservation is therefore necessary to retain the event-specific reconstruction learned by MoRF.

The symmetric Bures map attains a nine-domain mean ACE of 0.0225, compared with 0.0270 for the Cholesky map. It is positive definite, maps the Gaussian reference covariance exactly to its scaled counterpart, and minimizes quadratic displacement between the same-center Gaussian references.

Together, these ablations support mean-preserving Bures transport as the AST map and show that direct high-dimensional covariance estimation does not improve calibration with the available history.

\subsubsection{Effect of Truncation Correction on Identity Gating}
\label{sec:gate-ablation}

\begin{figure}[htbp]
    \centering
    \includegraphics[width=\textwidth]{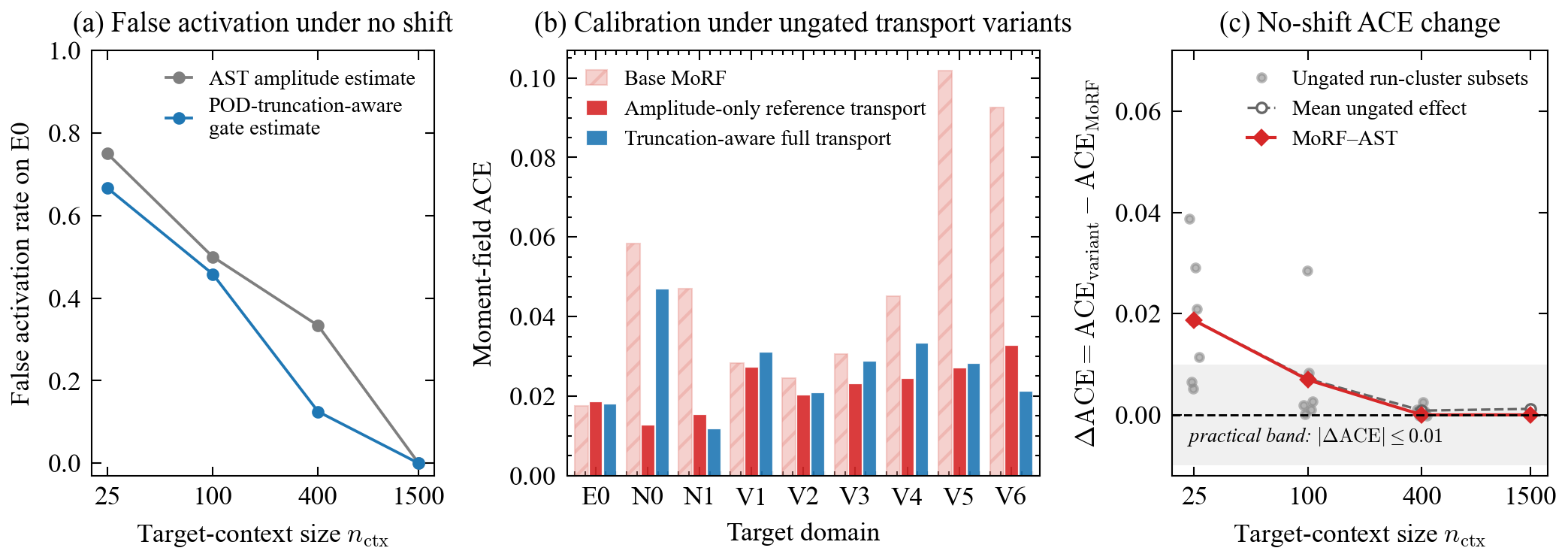}
    \caption{Truncation-aware identity gating under the no-shift control}
    \label{fig:gate-safety}
\end{figure}

To test whether unresolved POD variance causes false transport, we compare the ordinary and truncation-aware scale estimates on repeated E0 subsets (Fig.~\ref{fig:gate-safety}). The truncation-aware estimate activates less often, while both estimates return identity for the full context. Accounting for unresolved sensor-space energy therefore protects the frozen posterior from unnecessary adaptation.

Carrying the same truncation correction into the transported covariance raises aggregate ACE in most domains. The final design therefore assigns \(\widehat\alpha_g\) to the gate and \(\widehat\alpha_v\) to the retained-space spread map. This separation preserves a conservative identity decision with moderate histories without injecting unresolved POD energy into transport.

\subsection{Sensitivity to Deployment Information and Measurement Assumptions}
\label{sec:deployment-sensitivity}

\subsubsection{Sensitivity to Historical Context Size}
\label{sec:context-size}

\begin{figure}[htbp]
    \centering
    \includegraphics[width=0.5\columnwidth]{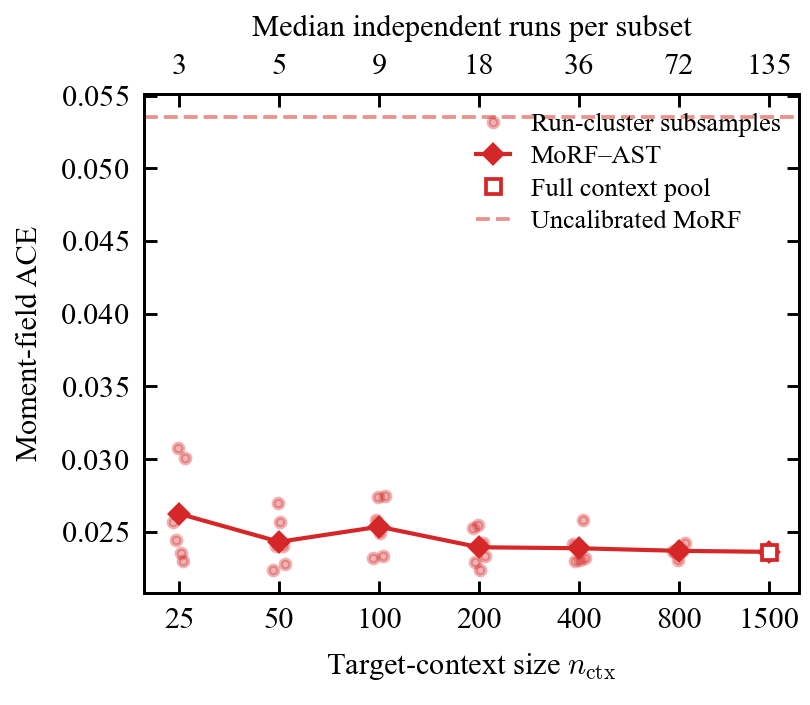}
    \caption{Historical-context sensitivity of MoRF-AST}
    \label{fig:context-size}
\end{figure}

To determine the history needed for stable scale estimation, we evaluate context budgets from 25 to 800 events using six independently assembled run clusters at each partial budget (Fig.~\ref{fig:context-size}). The full set of 1,500 events is evaluated once. Budgets are assembled from complete simulated traffic runs, so response snapshots within a run are not treated as independent operational realizations.

Even the smallest context budget improves the shifted-domain calibration relative to unadapted MoRF, although its repetition-to-repetition variability is appreciable. The eight-domain mean ACE is 0.0262 at the nominal 25-event budget, 0.0239 at 400 events, and 0.0236 with the full 1,500-event context. The result therefore stabilizes by the nominal 400-event budget, corresponding to a median of approximately 36 independent traffic runs and providing a practical operating point for scale estimation.

\subsubsection{Sensitivity to Sensor Coverage}
\label{sec:sensor-count}

\begin{figure}[htbp]
    \centering
    \includegraphics[width=0.5\columnwidth]{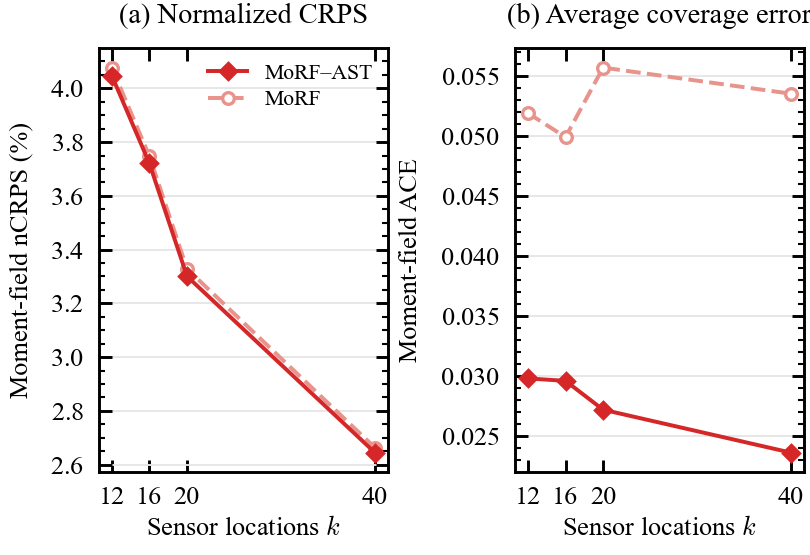}
    \caption{Sensor-count sensitivity of MoRF and MoRF-AST}
    \label{fig:sensor-count}
\end{figure}

To assess how sensor coverage affects reconstruction and calibration, we evaluate nested layouts with \(k=12,16,20,\) and \(40\) locations from the same farthest point ordering (Fig.~\ref{fig:sensor-count}). Every location supplies four channels, so the number of scalar observations is \(4k\) for \(d=192\) retained coefficients. Source channel scales, the POD basis, the master noise realization, and a common scoring mask that excludes all 40 master sensor positions were fixed. Each sensor count used its own observation operator, Gaussian reference, and trained flow checkpoint. The nested layouts therefore change sensing information while retaining a common spatial design and evaluation region.

Increasing sensor coverage steadily lowers nCRPS because the query constrains more of the retained state. From \(k=12\) to \(k=40\), the MoRF nCRPS decreases from \(4.08\%\) to \(2.66\%\), while the MoRF-AST value decreases from \(4.05\%\) to \(2.64\%\). At \(k=12\), AST reduces ACE from 0.0519 to 0.0298. At \(k=40\), it reduces ACE from 0.0535 to 0.0236, and the same ordering holds at \(k=16\) and \(k=20\). Base MoRF calibration does not improve monotonically with sensor count. Additional query information alone therefore does not remove the operating-scale spread bias. Using common base samples within each layout confirms that AST improves calibration across all tested sensor counts through spread transport rather than posterior sampling variation.

\subsubsection{Sensitivity to Observation Noise Misspecification}
\label{sec:noise-sensitivity}

\begin{figure}[htbp]
    \centering
    \includegraphics[width=0.5\columnwidth]{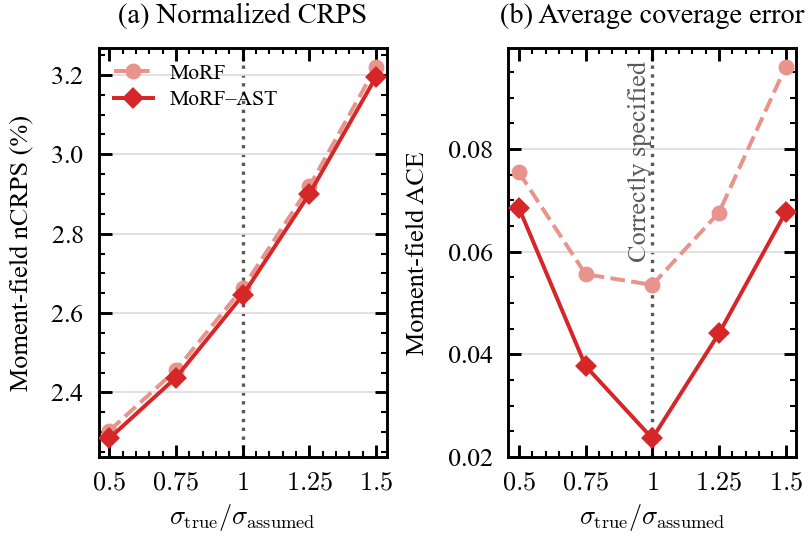}
    \caption{Observation-noise sensitivity of MoRF and MoRF-AST}
    \label{fig:noise-sensitivity}
\end{figure}

To assess robustness to noise misspecification, we hold the assumed covariance \(R=0.05^2I\) fixed while scaling the true noise applied to both query and context observations (Fig.~\ref{fig:noise-sensitivity}). The same clean events and seeded independent Gaussian draws are used for every true to assumed noise standard deviation ratio from 0.5 to 1.5. This construction isolates the imposed noise magnitude from event and noise realization changes.

Increasing the true noise raises nCRPS because the measurements carry less event-specific information. Calibration is best when the assumed and true noise levels agree and degrades when the model either overstates or understates measurement uncertainty. MoRF-AST nevertheless retains lower nCRPS and ACE than MoRF at every tested ratio. The adaptation advantage remains robust over true-to-assumed noise ratios of 0.5--1.5, while accurate noise characterization remains important for the scale fit.

\subsection{Evaluation of Recurrent-State Context Aggregation}
\label{sec:recurrent-context}

\begin{figure}[htbp]
    \centering
    \includegraphics[width=\textwidth]{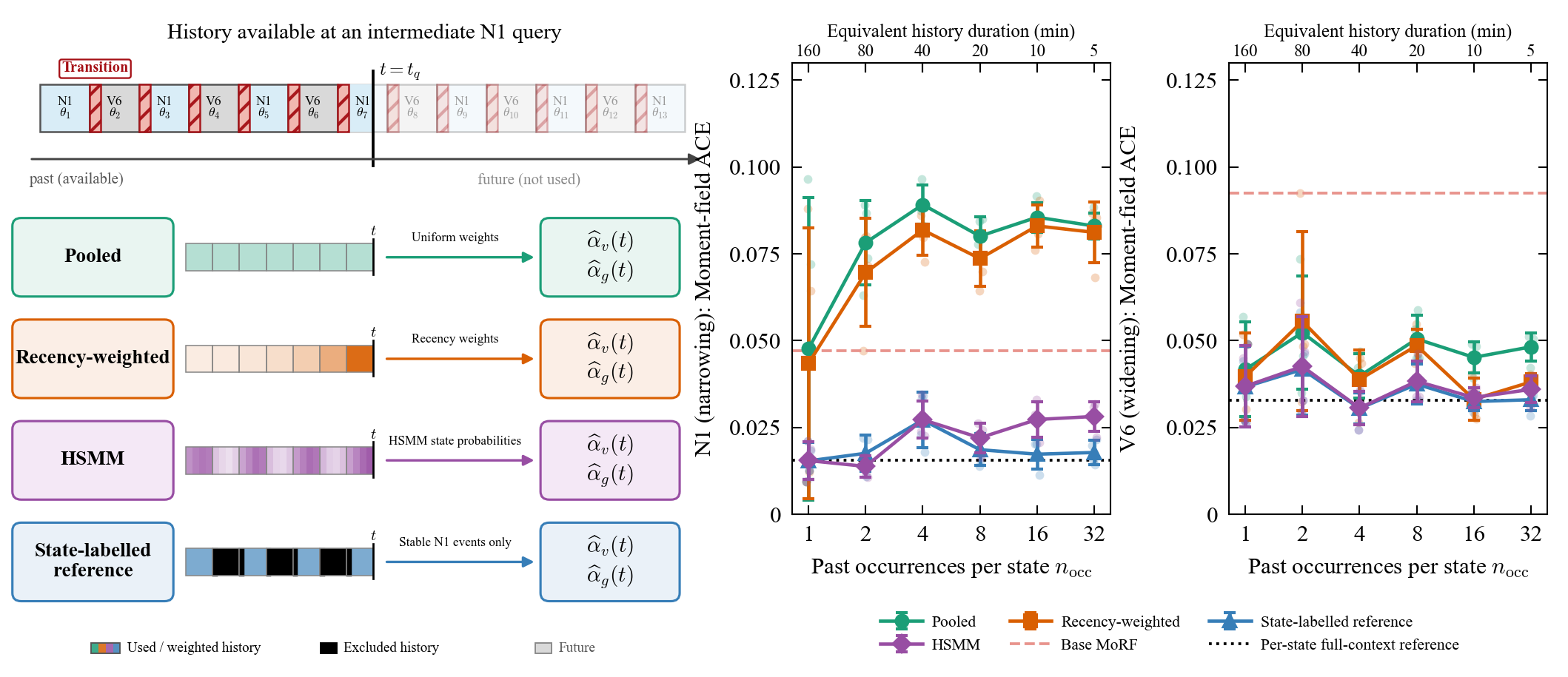}
    \caption{Context aggregation under recurrent operating states}
    \label{fig:recurrent-context}
\end{figure}

To support the practical use of AST, we make a preliminary comparison of several ways to select historical observations that represent the current operating condition. We construct 320-minute histories in which N1 and V6 alternate (Fig.~\ref{fig:recurrent-context}). For each evaluation state, \(n_{\mathrm{occ}}=1,2,4,8,16,\) or \(32\) past occurrences correspond to individual dwell times from 160 to 5 minutes. Moving right in the figure therefore represents more frequent switching. Four seeded histories are evaluated at each setting. Simulator state labels select the final stable evaluation point and construct a privileged reference, but they are not supplied to the observation-based history models.

\FloatBarrier

Four ways of assembling history are compared with the same AST scale objective. Pooled history gives uniform weight to every available observation before the query. Recency weighting favors recent observations. The observation-only hidden semi-Markov model (HSMM) \cite{yu2010hsmm} uses 60-second bins of sensor root-mean-square response and availability, then weights each past frame by its posterior probability of sharing the anonymous query state. This weighting can recover noncontiguous but state-compatible periods. If the inferred segments are too short, it returns to uniform pooling. The state-labeled case uses known simulator states and serves only as a privileged reference. No strategy accesses observations after the query time.

For both recurrent states, HSMM weighting approaches the calibration obtained from a full state-specific context while remaining well below base MoRF. It improves on pooled history in every tested state-frequency combination and on recency weighting in nearly every combination. The only reversal is negligible relative to the variation across histories. These trends show that matching the current state is more important than treating all past observations equally.

The state-dependent response of pooled history shows that indiscriminate aggregation mixes relevant and irrelevant observations. Observation-based state weighting instead reuses noncontiguous periods that match the current operating state, which preserves a more coherent scale estimate as switching becomes frequent. The tested histories support this mechanism in the recurrent night and congested setting, while broader state structures require separate evaluation.

Overall, adaptation is governed primarily by the relevance of operating history: moderate contexts stabilize the scale estimate, denser sensing improves accuracy but does not remove spread bias, and observation-based state weighting preserves calibration when conditions recur.

\section{Discussion}
\label{sec:discussion}

Across the numerical operating domains, changes in traffic composition and operating conditions were neither model inputs nor identified as individual parameters from sparse observations. Their effects propagated through the loading process and structural mapping and appeared in the second moment of sensor responses about the fixed source-domain mean \(b\), \(\mathbb E[(y-b)(y-b)^{\mathsf T}]\). AST uses the inferred scale component to adjust posterior spread, capturing an overall response-scale shift rather than specific traffic parameters. Similar signatures may arise from wind loads on tall buildings, wave loads on offshore platforms, or changes in rotating-machinery speed and load, suggesting applications of AST beyond traffic-induced response reconstruction.

With limited history, fitting a complex target-domain distribution is not necessarily preferable to estimating a scalar. For every layout with \(p<d\), distinct modal covariances can yield the same sensor covariance, preventing identification of a general high-dimensional covariance. AST instead estimates an effective scale \(\alpha\), reducing unsupported degrees of freedom without claiming uniqueness or optimality. The Gaussian conditional covariance and Bures map combine this scalar with sensor geometry to produce direction-dependent modal adjustments, a construction supported by the covariance-map ablation in Fig.~\ref{fig:map-ablation}. AST thus transfers the observable scale change to the full-field posterior through the existing conditional structure.

AST modifies posterior spread without correcting an erroneous posterior center. The base model must retain the main conditional response structure, and the imposed expansion or contraction must match its dispersion bias, as shown in Fig.~\ref{fig:directional-compatibility}. AST is therefore not a general calibrator, and its role also diminishes as observations become more informative. As sensor information in every retained direction grows without bound, the posterior concentrates and the Bures map approaches the identity, provided that \(\widehat\alpha_v\) remains finite and the standardized residual second moment remains bounded. Probabilistic virtual sensing then approaches point estimation within the retained subspace, leaving little spread for AST to adjust. Conversely, when observations are insufficiently informative and the reconstruction must rely on the prior, AST can provide effective corrections along those directions.

MoRF-AST is best suited when the source representation and observation relation remain valid, the operating condition remains within source-domain coverage, and \(R+\alpha S_y\) adequately represents the mean-referenced operating sensor second moment. Because this moment also reflects operating-mean shifts and noise misspecification, \(\widehat\alpha\) is only an effective response scale. New response modes, structural changes, pronounced anisotropic shifts, or sensor faults may exceed both AST and the base model. They may also invalidate fixed-source assumptions shared by other data-driven models, making them broader out-of-distribution challenges. Although MoRF-AST is better calibrated than the evaluated baselines for marginal prediction intervals, this does not establish accurate spatially joint extremes, exceedance probabilities, or tail risk. Accordingly, the AST-adjusted posterior may trigger dedicated tail analysis rather than replace high-fidelity evaluation of extreme responses and tail risk.

\section{Conclusion}
\label{sec:conclusion}

This study proposes MoRF-AST for probabilistic full-field reconstruction from sparse and noisy measurements under changing operating conditions. Evaluation covers a bridge-deck benchmark across shifted traffic domains, with the main conclusions summarized below.

(1) MoRF combines an analytic Gaussian reference with flow matching in posterior-whitened residual coordinates. The reference focuses learning on residual structure beyond the Gaussian approximation. With 8,000 training events, MoRF achieves an NRMSE of 7.20\%, less than half the corresponding NRMSEs of the two direct conditional flows.

(2) AST estimates an effective response scale from sensor-only operating history. It applies a truncation-aware identity decision and performs mean-preserving Bures transport when required. Across eight shifted domains, it reduces MoRF's mean ACE from 0.0535 to 0.0236, and the identity gate correctly avoids unnecessary adaptation. Further ablation and sensitivity studies support the MoRF-AST design choices across the evaluated deployment settings.

(3) Transport improves calibration only when its direction matches the base posterior's spread bias. MoRF's bias aligns with the required action in both the contraction and expansion regimes, whereas the biases of the other baselines do not. MoRF thus gives AST a calibrated base posterior with correctable spread bias under domain shifts.

Overall, this study offers design insights for diffusion- and flow-based generative methods in engineering probabilistic virtual sensing. It also addresses posterior calibration under changing operating conditions, a significant problem that has received limited attention. By enabling observation-only calibration during deployment, MoRF-AST supports long-term probabilistic full-field reconstruction and safer operation under changing conditions.
 
\section*{CRediT authorship contribution statement}

Wingho Feng: Data curation, Formal analysis, Methodology, Validation, Writing--original draft, Writing--review \& editing, Visualization. Quanwang Li: Supervision, Project administration, Resources. Ming Zhong: Project administration, Resources. Jingyu Yang: Project administration, Resources. Chen Wang: Conceptualization, Methodology, Writing--review \& editing.

\section*{Funding}

This work was supported by the National Key R\&D Program of China (2025YFF0518604) and the Beijing Nova Program (202604841290).

\section*{Data availability}

The numerical benchmark dataset and source code are available at \url{https://github.com/winghofeng/MoRF-AST}.

\bibliographystyle{unsrtnat}
\bibliography{references}

\end{document}